\documentclass[]{spie}  %>>> use for US letter paper
\usepackage{amsmath,amsfonts,amssymb}
\usepackage{graphicx}
\usepackage[colorlinks=true, allcolors=blue]{hyperref}
\usepackage{bm}

\usepackage{algpseudocode}
\usepackage{algorithm}
\usepackage{xcolor}
\usepackage{subfig}
\usepackage{comment}
\usepackage{booktabs}

\DeclareMathOperator*{\argmax}{arg\,max}

\newcommand{\R}{\mathbb{R}}

\title{Pseudo Closed-Loop Bootstrapping for Reinforcement Learning Based AO Control}

\author[1,2]{Eemil Halonen}
\author[1]{Jalo Nousiainen}
\author[1]{Markus Kasper}
\author[2]{Nuutti Hyvönen}
\affil[1]{European Southern Observatory, Karl-Schwarzschild Strasse 2, Garching, Germany}
\affil[2]{Aalto University, Otakaari 1, Espoo, Finland}

\authorinfo{Further author information: \\E. Halonen: E-mail: eemil.halonen@eso.org, \\  J.Nousiainen: E-mail: jalo.nousiainen@eso.org}

\begin{document} 
\maketitle

\begin{abstract}
High contrast imaging with ground-based telescopes requires extremely precise wavefront control. The performance of the controller ultimately depends on the quality and efficiency of the wavefront sensing technique, which has led to the development of ever more sensitive wavefront sensors (WFS). However, the increase in sensitivity comes at the cost of greater nonlinearity, which poses challenges for conventional linear reconstructors and controllers.

Reinforcement Learning (RL) is a branch of machine learning in which control policies are learned by interacting with the environment. RL has recently attracted interest in the field of XAO, and previous studies have demonstrated that a model-based RL variant, the Policy Optimization for Adaptive Optics (PO4AO), can effectively compensate for temporal delays, misregistration errors, and moderate WFS nonlinearities. PO4AO learns an initial control strategy by collecting closed-loop data from the integrator controller with a linear reconstructor. However, in cases of highly non-linear WFS and challenging conditions, closing the loop with the integrator is difficult, and high-fidelity data cannot be collected. Consequently, learning robust control with PO4AO is challenging. 

We propose a robust learning strategy in which PO4AO is pre-calibrated using an internal light source and a DM, enabling direct on-sky closed-loop operation without a stable integrator controller for bootstrapping the initial models. 
\end{abstract}

\newpage
\section{INTRODUCTION}
\label{sec:intro}  % \label{} allows reference to this section
Imaging Earth-like exoplanets around nearby stars is one of the primary science goals for the upcoming generation of extremely large telescopes. Achieving this goal requires further development of high-contrast imaging (HCI) techniques. Current HCI instruments on ground-based telescopes — such as SPHERE (Spectro-Polarimetric High-contrast Exoplanet REsearch \cite{fusco2006high, 2019A&A...631A.155B}), MagAO-X (Magellan Adaptive Optics eXtreme system \cite{males2018magao}), the Gemini Planet Imager \cite{macintosh2014first}, and SCExAO (Subaru Coronagraphic Extreme Adaptive Optics \cite{jovanovic2015subaru}) — combine extreme adaptive optics (XAO) with coronagraphy to suppress stellar diffraction. A fundamental limitation remains the residual wavefront error introduced by the XAO system, compounded by non-common-path errors. These uncorrected phase errors introduce speckles (i.e., residual starlight) in the focal plane that are indistinguishable from a faint companion at similar angular separations.

The XAO residual errors are mainly due to time-delay errors, arising from the inherent time lag between the wavefront measurement and the subsequent control signal to the XAO deformable mirror (DM). This has spurred the development of highly sensitive wavefront sensors, such as the Zernike wavefront sensor (ZWFS, \cite{zernike1934diffraction, bloemhof2003phase}), its variants (e.g., PIAAI-ZWFS\cite{haffert2023reaching}, vector ZWFS \cite{doelman2018simultaneous}, Z2WFS \cite{chambouleyron2021variation}), the bivortex WFS \cite{chambouleyron2024coronagraph}, the Bi–O edge WFS \cite{verinaud2024bi} (sharp edge), and more traditional, the non-modulated pyramid wavefront sensor (PWFS \cite{ragazzoni1996pupil}). These sensors achieve high sensitivity by operating in a regime where the wavefront-to-intensity mapping is steep but inherently nonlinear, making it hard to obtain stable closed-loop performance on full turbulence and more suited, for example, to running a second AO stage of a cascaded system \cite{n2024cascade, 2022JATIS...8a9001C}.

The mismatch between the assumed linear response and the true sensor behavior is the central challenge motivating non-linear reconstruction and control algorithms \cite{hutterer2018nonlinear, hutterer2019real, chambouleyron2024using, frazin2018efficient, haffert2024into}, including data-driven methods \cite{Landman:20,  wong2023nonlinear, archinuk2023mitigating, weinberger2024transformer}. Machine learning offers a route to nonlinear AO control using data, bypassing the need for detailed WFS modeling and providing a highly parallelizable, high-speed reconstruction operator that utilizes GPUs. This kind of Neural network (NN) reconstructor has been demonstrated in closed-loop laboratory conditions for the unmodulated PWFS on MagAO-X \cite{landman2024making, landman2025making}, enabling non-modulated operation of the PWFS under various atmospheric conditions at loop speeds $>$ 1KHz.

Reinforcement learning (RL) extends this further by learning predictive control in parallel with reconstruction through direct interaction with the AO environment, enabling online adaptation of the reconstruction and predictive control to prevailing optical gains (seeing) and wind conditions. Policy Optimization for Adaptive Optics (PO4AO \cite{nousiainen2022toward}) has been demonstrated to have stable performance with moderately linear WFSs in numerical simulations\cite{nousiainen2022toward}, lab set-ups\cite{nousiainen2024laboratory, nousiainen2022advances}, and on-sky\cite{nousiainen2026onskydemonstrationreinforcementlearning}. PO4AO learns an initial control strategy by collecting closed-loop data from an integrator controller with a linear reconstructor. However, under highly nonlinear WFS (e.g., nonmodulated PWFS or ZWFS) conditions or challenging atmospheric seeing, closing the loop with the integrator is difficult, and such data cannot be collected, making robust learning with PO4AO challenging. In this work, we propose a warmup calibration strategy in which the controller is pre-calibrated using an internal light source and a deformable mirror, enabling direct on-sky closed-loop operation without requiring a stable integrator controller. We benchmark both variants in COMPASS simulation across a range of atmospheric conditions, using the Strehl ratio and raw PSF contrast as the primary performance metric.
\newpage
\section{methods}

\subsection{PO4AO}
\label{sec:po4ao}
Policy Optimization for Adaptive Optics (PO4AO)\cite{nousiainen2024laboratory} is a two-model RL algorithm used for AO control. The two models are fully convolutional neural networks: a dynamics model and a policy model. 

The dynamics model \(\hat{p}_\omega\) models the AO control system. It predicts the next WFS reconstruction ${\bm{o}}_{t+1}$ from past telemetry data (past measurements and commands), that is,
\begin{equation*}
\tilde{\bm{o}}_{t+1} = \hat{p}_\omega(\bm{o}_t, \bm{o}_{t-1}, \dots, \bm{o}_{t-k},\ 
\bm{a}_t, \bm{a}_{t-1}, \dots, \bm{a}_{t-k}),
\end{equation*}
where, $\bm{a}$ are DM commands, and $\bm{o}$ reconstructed WFS measurements, i.e.,
\begin{equation}
    \bm{o}_{t+1} = C \Delta \bm{w}_t,
\end{equation}
with \(C\) being the reconstruction matrix and \(\Delta \bm{w}_t\) WFS measurement.

Similarly, the policy model $\pi_{\theta}$ is a convolutional NN, but it defines a nonlinear control law that maps the current telemetry history \(\bm{s}_t\) to a residual DM command \(\bm{a}_t\):
\begin{equation}
\bm{a}_t = \pi_{\theta}(\bm{s}_t) = \pi_{\theta}(\bm{o}_t, \bm{o}_{t-1}, \dots, \bm{o}_{t-k},\ 
                          \bm{a}_{t-1}, \dots, \bm{a}_{t-k}).
\end{equation}
The full command sent to the deformable mirror is then obtained by temporal accumulation:
\begin{equation}
\bm{v}_t = \alpha \bm{v}_{t-1} + \bm{a}_t,
\end{equation}
where \(\alpha \approx 0.99\) is the leakage factor.

The dynamics model $\hat{p}_\omega$ is trained via supervised learning on a continuously collected dataset $\mathcal{D}$ of closed-loop transitions. 

The policy $\pi_{\theta}$ is trained using the learned dynamics model on the same data, with starting states $\bm{s}$ sampled from $\mathcal{D}$. The policy then generates actions, and short imagined trajectories of length $H$ (the planning horizon) are simulated by unrolling the dynamics model. The reward is defined as the negative squared Euclidean norm of the predicted reconstruction:
\begin{equation}
\hat{r}_\omega(\bm{s}_t, \bm{a}_t) = -\|\tilde{\bm{o}}_{t+1}\|^2,
\end{equation}
where $\tilde{\bm{o}}_{t+1}$ is the output of the dynamics model prediction:
\begin{equation*}
\tilde{\bm{s}}_{t+1} = \hat{p}_\omega(\bm{s}_t, \bm{a}_t).
\end{equation*}
The policy is then optimized by maximizing the cumulative reward over the 
planning horizon:
\begin{equation}
\label{eq:optimization_of_hatr}
\argmax_\theta \sum_{{\bf s} \in \mathcal{D}} \sum_{t=1}^{H} 
\hat r_\omega(\tilde{\bm s}_t, \pi_\theta(\tilde{\bm s}_t)),
\end{equation}
where
\begin{equation*}
\tilde{\bm s}_1 = {\bm s} \quad \text{and} \quad 
\tilde{\bm s}_{t+1} = \hat p_\omega(\tilde{\bm s}_t, \pi_\theta(\tilde{\bm s}_t)).
\end{equation*}
Both models are trained in parallel with the control loop, which allows them to continuously adapt to atmospheric conditions.

%\subsection{PiPO4AO}
%Instead of utilizing the reconstructed observations $\tilde{\bm o}_{t+1}$, the pixel-based variant of PO4AO takes the raw WFS pixels directly as an input. This removes the reconstruction step and the usage of the command matrix $C_m$ completely.
\subsection{Robust training}

\subsubsection{Pseudo Closed-Loop Bootstrapping}
\label{sec:pclb}
Under highly nonlinear wavefront sensor (WFS) conditions, closing the AO loop with a classical integrator and linear reconstructor is often difficult. Although PO4AO itself does not rely on linearity assumptions, its training requires stable closed-loop data. If the integrator is unable to stabilize the loop, collecting usable trajectories for initialization becomes impossible.

To overcome this limitation, we propose an alternative calibration strategy, which we refer to as Pseudo Closed-Loop Calibration Bootstrapping (PCLB). Instead of relying on a closed-loop integrator, PCLB uses the deformable mirror to simulate the atmospheric turbulence and optimal integrator commands. Phase screen $\bm{\phi}_t$ is projected into DM actuator space: 
\begin{equation}
    \bm{v}_t^{\text{atm}} = P(\bm{\phi}_t),
    \label{eq:projection}
\end{equation}
where $P: \R^{N_{\phi}} \rightarrow \R^{N_{a}}$, $N_{\phi}$ size of the phase screen and $N_{a}$ the size of the DM. Since the projected phase screen is fully known, the ideal reconstruction at each step is simply its negation of uncorrected DM turbulence:
\begin{equation}
    \bm{e}_t = -(\bm{v}_t^{\text{atm}} + \bm{v}_{t-1}^{\text{correction}}).
\end{equation}
In the next step, the ideal reconstruction will be fed to the system with an integrator gain $g$, creating an ideal integrator, i.e., expert controller. The following DM correction is given by
\begin{equation}
    \bm{v}_t^{\text{correction}} = g\bm{e}_t + \alpha\bm{v}_{t-1}^{\text{correction}}.
\end{equation}
where $\alpha$ is the leakage factor defined in Section \ref{sec:po4ao}. This creates a controlled pseudo-closed-loop AO system in which the dynamics can be learned without a stable integrator acting on non-linear WFS measurements.

We further promote diversity in the data by adding normalized, scaled random noise to the expert command $\bm{e}_t$. The full expert control actions, i.e., residual command, $\bm{a}_t$, during the pseudo-closed loop data collection, are given by
\begin{equation}
    \bm{a}_t = g\,\bm{e}_t + \bm{\epsilon}_t, \quad 
    \bm{\epsilon}_t \sim \mathcal{N}(\bm{0}, \bm{1}).
\end{equation}
 
At each time step, we also record the observation $\bm o_t$, i.e., the reconstructed or preprocessed WFS measurement that relates the data to expert actions.  Each resulting transition, i.e., AO time step $(\bm{o}_t, \bm{a}_t, r_t, \bm{o}_{t+1})$ is stored in the warmup replay buffer $\mathcal{D_w}$, where $r_t$ denotes the reward. Warmup data in $\mathcal{D_w}$ is then used to pre-train the dynamics model and policy, yielding initial model parameters $(\theta^*, \omega^*)$, that are carried forward to on-policy training with DM turbulence or directly on-sky. The complete pseudo-closed-loop process is presented in Algorithm \ref{alg:warmup}.

\begin{algorithm}[H]
\caption{Warmup Data Collection for Robust Pretraining}
\label{alg:warmup}
\begin{algorithmic}[1]
\For{$k = 1$ to $N_w$}
    \State Sample atmospheric parameters: 
    $(r_0, v_{\text{wind}}, \varphi_{\text{wind}}, m) \sim \mathcal{U}$
    \State Reset environment with sampled atmosphere
    \State Initialise: $\bm{v}_0^{\text{correction}} \gets \bm{0}$, 
    $\bm{e}_0 \gets \bm{0}$
    \State Sample episode gain:
    $g \sim \mathcal{U}(0, 1)$
    \For{$t = 0$ to $L-1$}
        \State Generate phase screen and project: 
        $\bm{v}_t^{\text{atm}} \gets P(\bm{\phi}_t)$
        \State Compute expert correction: 
        $\bm{e}_t \gets -(\bm{v}_t^{\text{atm}} + 
        \bm{v}_{t-1}^{\text{correction}})$
        \State Form noisy warmup action:
        $\bm{a}_t \gets g\,\bm{e}_t + \bm{\epsilon}_t$, \quad
        $\bm{\epsilon}_t \sim \mathcal{N}(\bm{0}, \bm{1})$
        \State Apply leaky accumulation:
        $\bm{v}_t^{\text{correction}} \gets g\,\bm{e}_t + 
        \alpha\,\bm{v}_{t-1}^{\text{correction}}$
        \State Add atmospheric disturbance and apply total command:
        $\bm{c}_t \gets \bm{v}_t^{\text{correction}} + \bm{v}_t^{\text{atm}}$
        \State Propagate one AO step, observe $\bm{o}_{t+1}$
        \State Compute reward: $r_t \gets -\|\bm{o}_{t+1}\|^2$
        \State Store $(\bm{o}_t, \bm{a}_t, r_t, \bm{o}_{t+1})$ 
        in $\mathcal{D}_w$
    \EndFor
\EndFor
\State Pre-train $(\theta^*, \omega^*)$ on $\mathcal{D}_w$
\end{algorithmic}
\end{algorithm}

\subsubsection{On-sky adaption}
During the PCLB the atmospheric conditions (i.e., wind directions, seeing) are drawn randomly and, hence the pretrained model is optimized to obtain stable performance in every possible condition encountered in the observation site. However, to reach optimal performance PO4AO need also to adapt to prevailing conditions. This is done by training the models in parallel to training closed loop operation; detailed in this section.

After PCLB pre-training, the policy $\pi_{\theta}$ and the dynamics model $\hat{p}_{\omega}$ are deployed directly for on-sky operation without requiring a classical integrator. To enable continuous adaptation to varying atmospheric conditions, both models are further trained online during closed-loop operation. On-sky conditions are simulated in COMPASS, which creates realistic Kolmogorov turbulence that matches typical observing conditions based on our predefined parameters. 

\begin{algorithm}[H]
\caption{On-sky Adaptation and Continuous Learning, i.e., classic PO4AO without warm-up}
\label{alg:onsky_adaptation}
\begin{algorithmic}[1]
\State Load pre-trained $\pi_{\theta^*}$, $\hat{p}_{\omega^*}$
\State Initialise online buffer $\mathcal{D}_{\text{on}} \gets \emptyset$
\State Close AO loop with $\pi_{\theta^*}$
\For{each episode $i$}
    \For{each timestep $t$}
        \State Observe state $\bm{s}_t$
        \State Compute action: $\bm{a}_t \gets \pi_\theta(\bm{s}_t)$
        \State Apply DM command: 
        $\bm{v}_t \gets \alpha \bm{v}_{t-1} + \bm{a}_t$
        \State Observe next state $\bm{s}_{t+1}$, compute 
        $r_t \gets -\|\bm{o}_{t+1}\|^2$
        \State Store $(\bm{s}_t, \bm{a}_t, r_t, \bm{s}_{t+1})$ 
        in $\mathcal{D}_{\text{on}}$
    \EndFor
    \State Sample $\mathcal{B} \sim \mathcal{D}_{\text{on}} \cup 
    \mathcal{D}_w$
    \State Update $\hat{p}_\omega$ by minimising $\mathcal{L}_\omega$ 
    over $\mathcal{B}$ \hfill (Eq.~\ref{eq:dynamics_loss})
    \State Update $\pi_\theta$ by maximising $\hat{r}_\omega$ 
    over $\mathcal{B}$ \hfill (Eq.~\ref{eq:policy_opt})
\EndFor
\end{algorithmic}
\end{algorithm}

As described in Algorithm \ref{alg:onsky_adaptation}, at the beginning of an on-sky simulation run, the pre-trained models are loaded and the AO loop is closed immediately. The system then collects closed-loop transitions $(\bm{s}_t, \bm{a}_t, r_t, \bm{s}_{t+1})$ into an online replay buffer $\mathcal{D}_{\text{on}}$. Mini-batches that are used for model updates are sampled from the union of the online buffer and the warmup buffer: 
\begin{equation}
    \mathcal{B} \sim \mathcal{D}_{\text{on}} \cup \mathcal{D}_w,
\end{equation}
where $\mathcal{D}_{w}$ is the warmup buffer defined in Section \ref{sec:pclb}. 
Mixing samples from both replay buffers stabilizes the learning, by reducing overfitting to recent atmospheric conditions, avoiding losing memory of noisy "bad" actions, and to ensure mini-batches remain diverse and temporally uncorrelated.

The dynamics model is updated by minimizing the mean squared prediction error:
\begin{equation}
    \mathcal{L}_{\omega} = \mathbb{E}_{\mathcal{B}} 
    \left[ \| \bm{o}_{t+1} - \hat{p}_{\omega}(\bm{s}_t, \bm{a}_t) \|^2 
    \right],
    \label{eq:dynamics_loss}
\end{equation}
and the policy is optimized by maximizing the cumulative reward over the planning horizon $H$:
\begin{equation}
    \argmax_\theta \sum_{\bm{s} \in \mathcal{B}} \sum_{t=1}^{H} 
    \hat{r}_\omega(\tilde{\bm{s}}_t, \pi_\theta(\tilde{\bm{s}}_t)),
    \label{eq:policy_opt}
\end{equation}
where the imagined trajectory is defined recursively $\tilde{\bm{s}}_{t+1} = \hat{p}_\omega(\tilde{\bm{s}}_t, 
\pi_\theta(\tilde{\bm{s}}_t))$ and the stepwise reward is given by $\hat{r}_\omega(\bm{s}_t, \bm{a}_t) = -\|\tilde{\bm{o}}_{t+1}\|^2$. Both models are updated in parallel throughout the observation.

\newpage
\section{Results}
\label{sec:results}
This section addresses the parameters used for numerical simulations and introduces results. We evaluate the proposed PCLB-PO4AO (hereafter PCLB) against the original PO4AO and a classical integrator (integrator gain tuned for each case separately), assuming linear reconstruction. Numerical simulations are conducted using COMPASS simulator. We simulate an 8\,m telescope, with a non-modulated PWFS and a 41x41 DM, and closed the loop in variety of different seeing conditions with all the above controllers.  All parameters used for the simulation are presented in Table \ref{tab:simulation_parameters}.

\begin{table}[htbp]
\centering
\caption{Simulation and Training Parameters}
\label{tab:simulation_parameters}
\begin{tabular}{lll}
\toprule
\textbf{Parameter} & \textbf{Value} & \textbf{Unit} \\
\midrule
\multicolumn{3}{l}{\textbf{On-sky / Simulation}} \\
\midrule
Number of atmospheric layers & 3 & -- \\
Effective wind speed & 20.35 & m/s \\
Average wind direction & 120 & $^\circ$ \\
Star magnitude & 4 & mag \\
Fried parameter & 5-20& cm\\
\midrule
\multicolumn{3}{l}{\textbf{Telescope}} \\
\midrule
Type & VLT & -- \\
Diameter & 8 & m \\
Pupil diameter & 480 & pixels \\
Central obstruction ratio & 0.14 & -- \\
\midrule
\multicolumn{3}{l}{\textbf{AO System}} \\
\midrule
DM & $41 \times 41$ & actuators \\
WFS wavelength & 850 & nm \\
\bottomrule
\end{tabular}
\end{table}

In general, the original PO4AO works really robustly with non-modulated PWFS in numerical simulations, so we had to extend seeing conditions worse than is expected on good observing sites. We simulated a strong wind, $\tau = 20.35$ms, and  tested all algorithms (original PO4AO, PCLB, integrator) with  $r_0 \in [0.05$, $0.20]$\/. We note that PCLB was not trained/bootsrapped for any specific seeing conditions, wind direction, or guide star magnitude, but the same exact model is applied for every operation. 

\begin{figure}[H]
    \centering
    \includegraphics[width=0.5\linewidth]{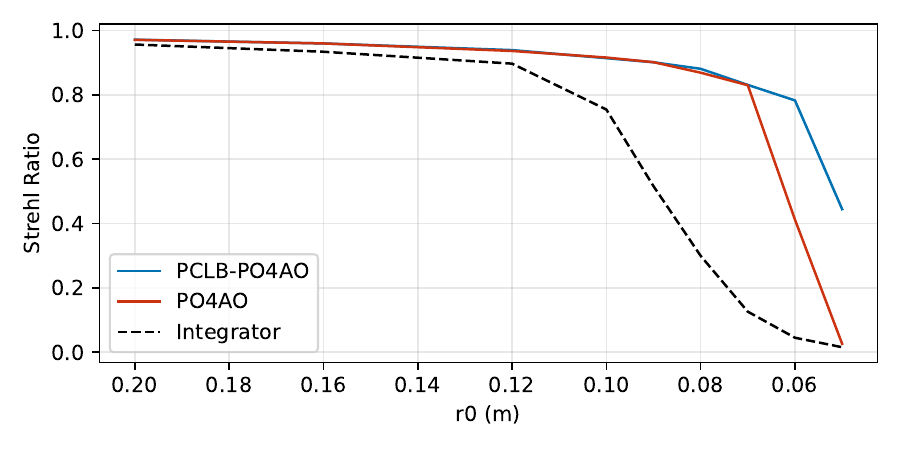}
    \caption{Converged Strehl ratio as a function of $r_0$. PCLB maintains stable performance into poor seeing conditions where standard PO4AO and the integrator degrade significantly.}
    \label{fig:comparison}
\end{figure}

Performance is measured by the Strehl ratio and total reward (negative normalized WFS variance) per iteration during the exposure. Figure \ref{fig:comparison} shows the final performance (after 60 iterations) of each method. 

One iteration here is 500 time steps, and during each iteration, the policy and dynamics are trained, and the control policy is replaced with updated parameters. We further examine the performance by recording PSF and calculated raw contrast as a function of angular separation $(\lambda/D)$ at convergence. Results are presented in three batches, each for a different seeing condition. 
 
Figure \ref{fig:comparison} shows the Strehl ratio across all tested $r_0$ values. Under good to moderate seeing ($r_0 \geq 0.12$\,m), all three methods achieve comparable performance. Below $r_0 = 0.12$\,m, the integrator and original PO4AO start to decrease in performance rapidly, while PCLB maintains a stable Strehl ratio down to $r_0 = 0.07$\,m before degrading under the worst conditions.
\begin{figure}[htbp]
    \centering
    \begin{minipage}{0.49\linewidth}
        \centering
        \includegraphics[width=\linewidth]{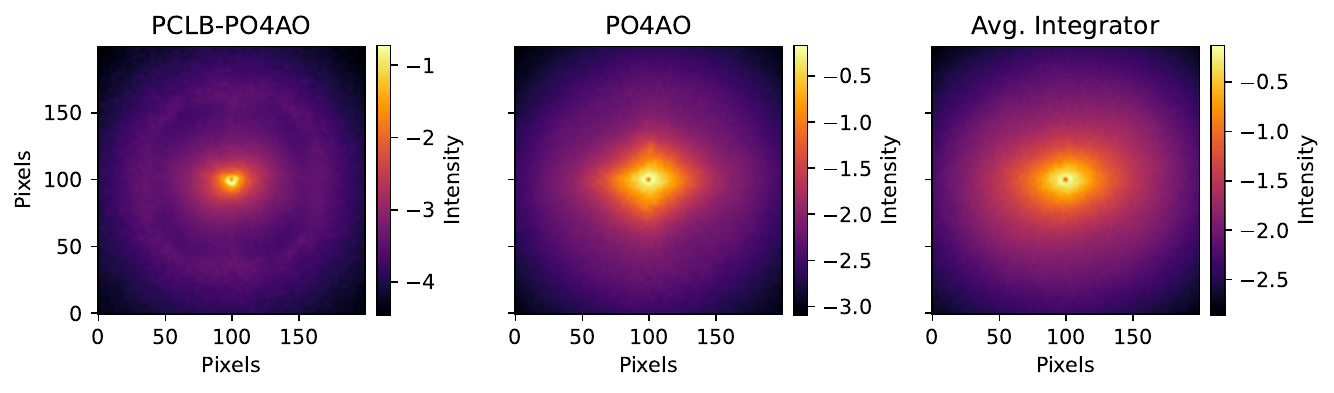}
        \includegraphics[width=\linewidth]{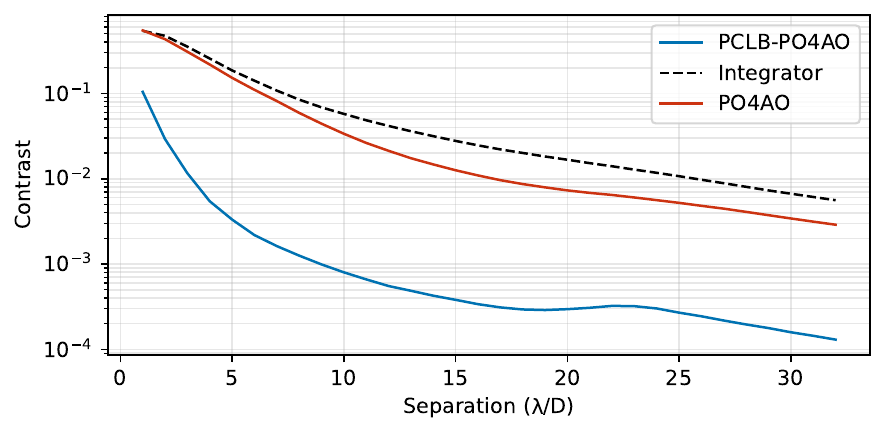}
    \end{minipage}
    \hfill
    \begin{minipage}{0.49\linewidth}
        \centering
        \includegraphics[width=\linewidth]{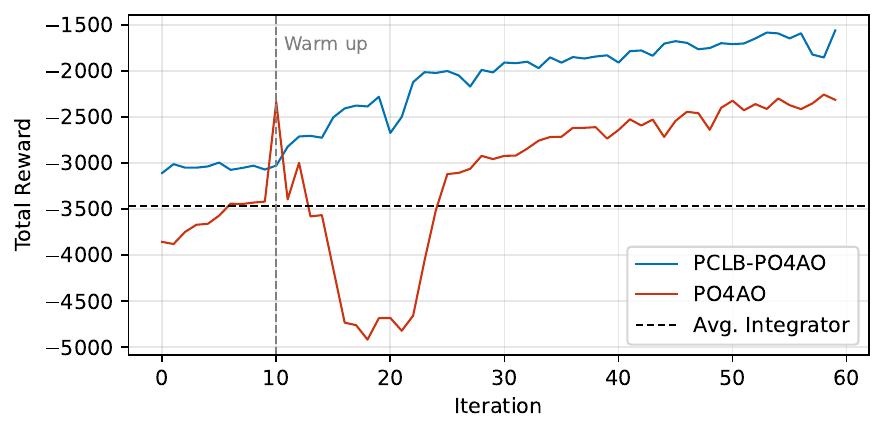}
        \includegraphics[width=\linewidth]{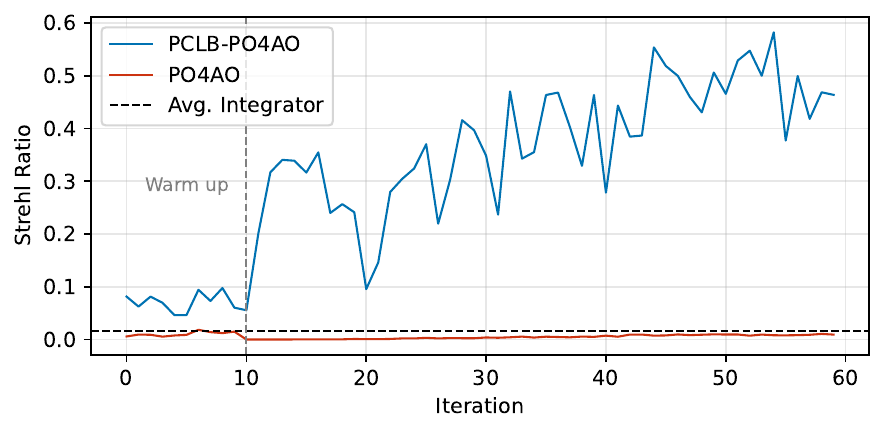}
    \end{minipage}
    \caption{Poor seeing ($r_0 = 0.05$\,m). Only PCLB achieves 
    a stable Strehl ratio and concentrated PSF.}
    \label{fig:combined_r005}
\end{figure}

\begin{figure}[htbp]
    \centering
    \begin{minipage}{0.49\linewidth}
        \centering
        \includegraphics[width=\linewidth]{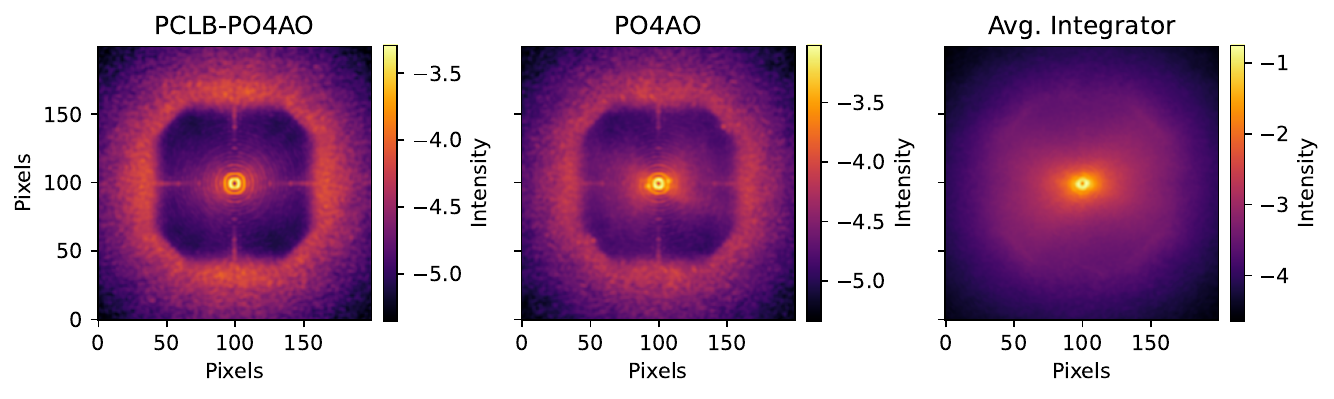}
        \includegraphics[width=\linewidth]{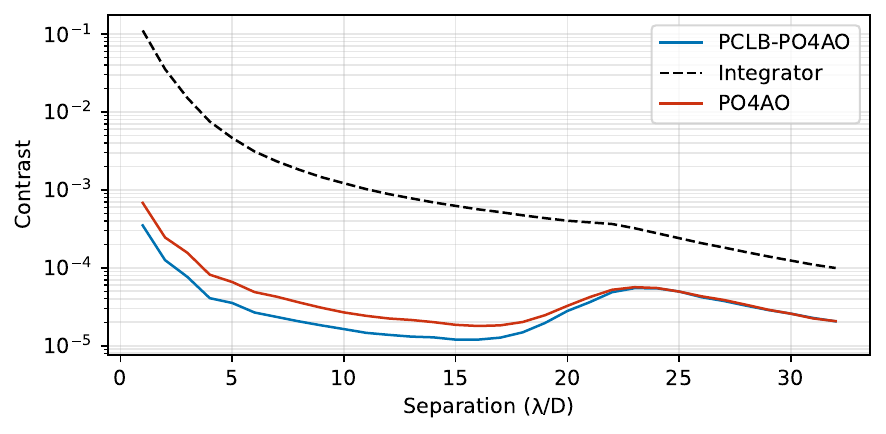}
    \end{minipage}
    \hfill
    \begin{minipage}{0.49\linewidth}
        \centering
        \includegraphics[width=\linewidth]{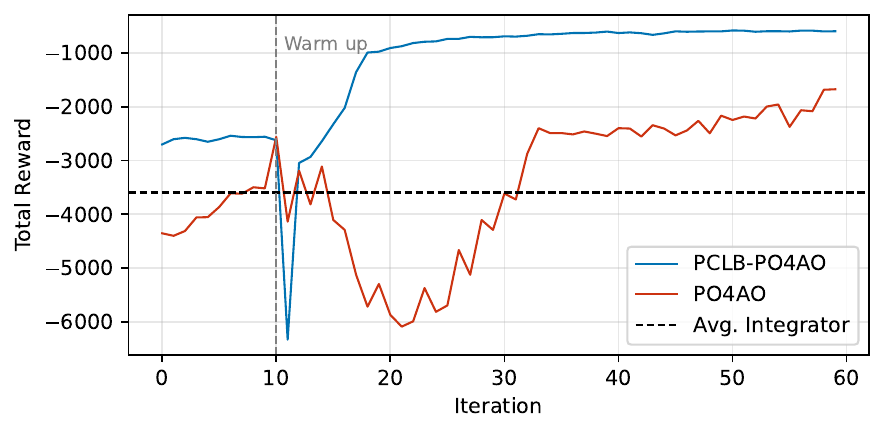}
        \includegraphics[width=\linewidth]{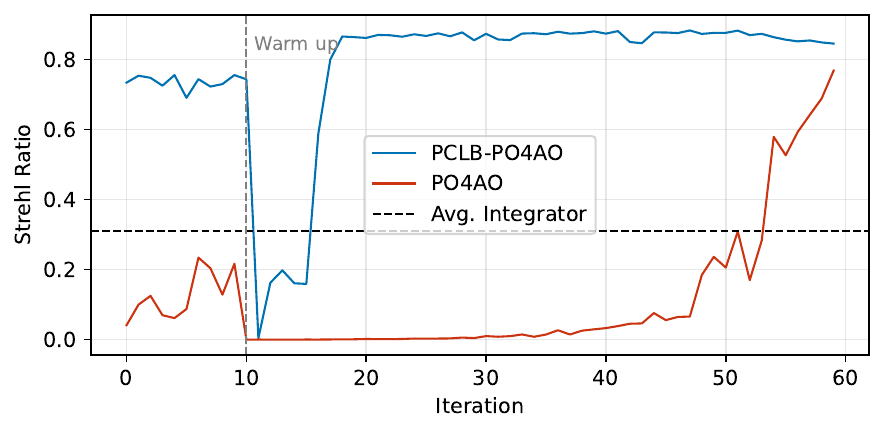}
    \end{minipage}
    \caption{Moderate seeing ($r_0 = 0.08$\,m). PCLB converges 
    after warmup while standard PO4AO and the integrator remain unstable.}
    \label{fig:combined_r008}
\end{figure}

\begin{figure}[htbp]
    \centering
    \begin{minipage}{0.49\linewidth}
        \centering
        \includegraphics[width=\linewidth]{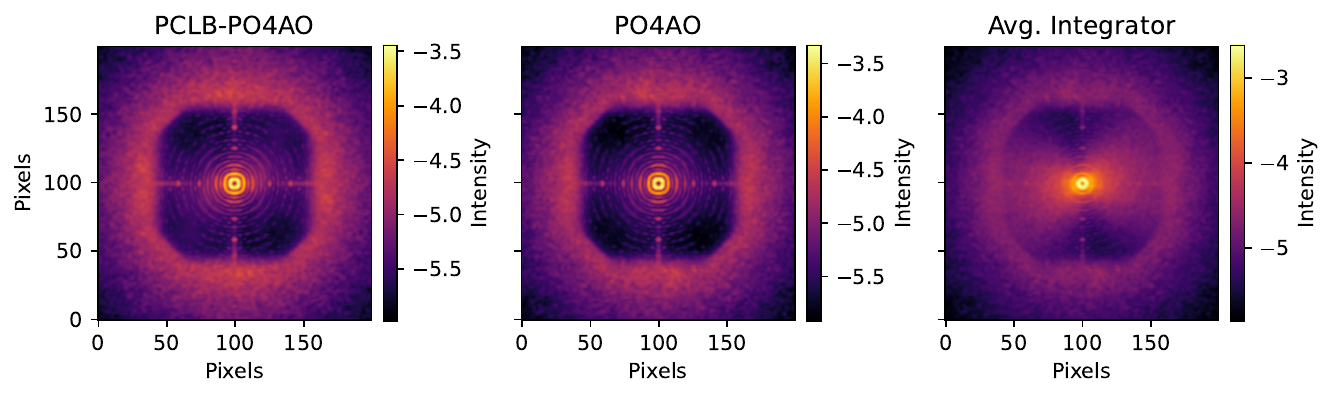}
        \includegraphics[width=\linewidth]{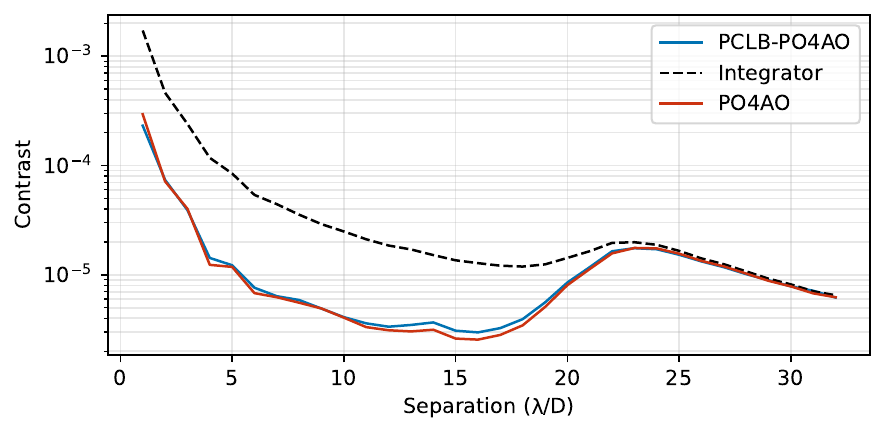}
    \end{minipage}
    \hfill
    \begin{minipage}{0.49\linewidth}
        \centering
        \includegraphics[width=\linewidth]{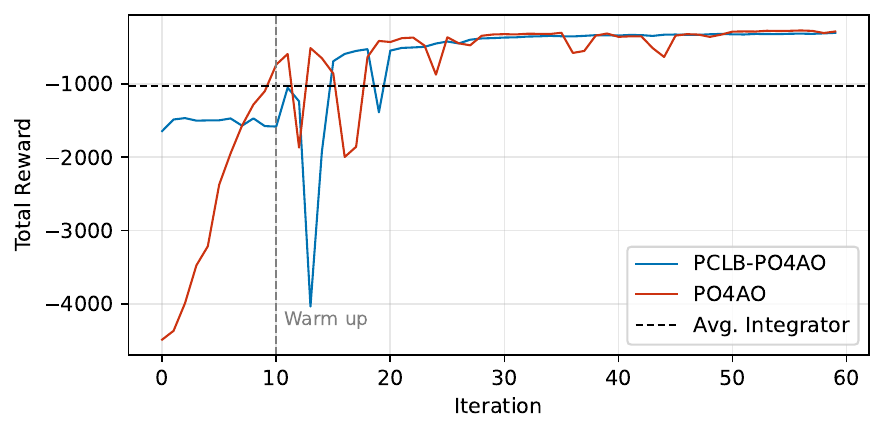}
        \includegraphics[width=\linewidth]{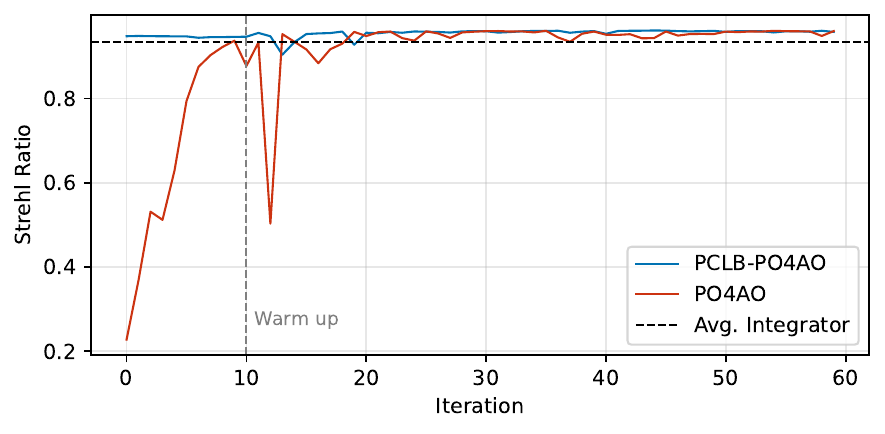}
    \end{minipage}
    \caption{Good seeing ($r_0 = 0.16$\,m). All methods converge; 
    PCLB and PO4AO outperform the integrator, in contrast, with 
    PCLB converging faster.}
    \label{fig:combined_r016}
\end{figure}

Under poor seeing ($r_0 = 0.05\,m$), both the integrator and the original PO4AO fail to converge, producing scattered PSFs, while PCLB maintains a concentrated PSF, stable Strehl ratio (Figure \ref{fig:combined_r005}), and good contrast. At moderate seeing ($r_0 = 0.09 \,m$), PCLB remains the only method achieving stable convergence (Figure \ref{fig:combined_r008}). Under good seeing ($r_0 =0.20\,m$), all methods converge, though the integrator produces worse contrast and original PO4AO shows transitory instability following the warm-up phase (Figure \ref{fig:combined_r016}).

\section{Discussion}
The results presented in \ref{sec:results} demonstrate that PCLB achieves stable performance across all tested seeing conditions, while standard PO4AO becomes unstable under difficult atmospheric conditions with the non-modulated PWFS. As seen in the contrast curve of Figure \ref{fig:combined_r005}, PCLB maintains competitive contrast even under poor seeing conditions. This is further supported by the long-exposure PSF images, where PCLB produces a relatively concentrated intensity distribution, while compared methods show significant spreading.

The Strehl ratio curves in Figure \ref{fig:combined_r005} further illustrate the advantage of being independent from the integrator: PCLB converges to a stable Strehl ratio throughout the observation, while standard PO4AO cannot be initialized under poor seeing conditions with a highly sensitive nonlinear WFS.

The benefit of the method extends to better seeing conditions as well. Under good seeing ($r_0 = 0.16$\,m), the achieved contrast is comparable across methods, as shown in Figure \ref{fig:combined_r016}. However, PCLB reaches stable performance immediately, whereas PO4AO requires iterations to learn from the integrator. The convergence speed difference is significant and could be meaningful in terms of on-sky operations where observation time is limited.

In conclusion, PCLB offers an alternative initialization strategy for RL-based AO control. Rather than relying on the classical integrator for warmup data collection, the proposed method generates initial training data by projecting simulated phase screens onto the DM. This creates a system of paired phase screen and DM action trajectories that bootstrap the RL training, which makes PO4AO independent of the classical integrator and its assumptions of linear reconstruction. Furthermore, the method allows the deployment of highly nonlinear but sensitive wavefront sensors. As the next step, we aim to validate these simulation results on the GHOST testbench.
\newpage
\bibliography{report} % bibliography data in report.bib
\bibliographystyle{spiebib} % makes bibtex use spiebib.bst

\end{document}